\documentclass[twoside]{article}
\usepackage[sc]{mathpazo}
\usepackage[T1]{fontenc} 
\usepackage{microtype}
\usepackage{multicol}
\usepackage{graphicx}
\usepackage{listings}
\usepackage{gensymb}
\usepackage{subcaption}
\usepackage{amssymb,amsfonts,amsmath}
\usepackage[hmarginratio=1:1,top=32mm,columnsep=20pt]{geometry} 
\usepackage{hyperref}
\usepackage{lettrine} 
\usepackage{textcomp}
\usepackage{abstract}
\usepackage{color,listings}
\definecolor{lightgray}{rgb}{.7,.7,.7}
\definecolor{white}{rgb}{1,1,1}
\newsavebox\lstbox

\usepackage{fancyhdr}
\title{\vspace{-12mm}\fontsize{24pt}{25pt}\selectfont\textbf{A Primordial Particle System in three dimensions}}
\author{
\large
\textsc{Thomas Schmickl\footnote{Corresponding author: \href{mailto:thomas.schmickl@uni-graz.at}{thomas.schmickl@uni-graz.at}}} \ and \textsc{Martin Stefanec}\\[2mm] \\ 
\small Artificial Life Lab of the Institute of Biology, Karl-Franzens-University Graz, Austria
 \\ \small Universit\"atsplatz 2, A-8010 Graz, Austria}
\date{\small submitted $27^{th}$ January 2019}

\begin{document}

\maketitle 

\thispagestyle{fancy} 

\begin{abstract}
This article describes the conversion of the two-dimensional Primordial Particle System into a three-dimensional model that exhibits comparable features. We present the transformed model here in the form of a pseudocode implementation and detail the modifications required for this conversion.
 \vspace{2mm}

\setlength{\parindent}{0pt}Keywords: Self-organization, Emergent pattern formation, Self-replication, Third Dimension, Artificial Life, Emergence of Life, Morphogenesis \vspace{2mm}

\end{abstract}

\section{Introduction}
 
A very simple model called PPS (\underline{P}rimordial \underline{P}article \underline{S}ystem) demonstrates that very simple rules (described by a simple motion law) can generate complex structures that exhibit many properties also found in life forms (growth, reproduction, physiology, nutrient cycle, life cycle, behaviour, information-processing, and finally also death) \cite{citation:1}, \cite{citation:2}. In addition populations of these individual emergent structures show similar behavior that is also found in animal populations (emergent ecosystems following density-dependent growth). This model differentiates itself from other, seemingly similar, self-propelled particle systems that show collective dynamics \cite{citation:3} by its simplicity. The original article and its associated demonstration movie triggered significant interest on the web (youtube\footnote{$https://www.youtube.com/watch?v=makaJpLvbow$}, reddit\footnote{$ https://www.reddit.com/search?q=\%22primordial\%20particle\%20system\%22$}, etc.) with several people starting to re-implement and modify it. A question often asked by the community is ``does it also work in 3D?''. Here our aim is to demonstrate that it is in fact trivial to extend the model into three dimensions and explain this with the amount of information that is needed to conduct this transition on the new spin-off code that was generated recently, in order to replicate and be able to check our observations.

\section{The mathematical model}

The original 2D model of the PPS assumes that a certain number of particles are initially positioned at randomized spots, with randomized orientation (uniform random distribution). For a 3D model implementation, these starting conditions are identical, just that the random positions are extended to \boldmath $<x_{i(t)}, y_{i(t)}, z_{i(t)}>$. \par \medskip
In our simulation we distributed 770 particles in a cubic space of 31 x 31 space units. Each particle i is orientated with two angles \boldmath $<\varphi^{xy_{i(t)}}, \varphi^{tilt_{i(t)}}>$ whereby \boldmath $\varphi^{xy}$ denotes the angle in the x-y plane (like in the original 2D version) and \boldmath $\varphi^{tilt}$ denotes the angle in the x-y plane (like in the original 2D version) and $\varphi^{tilt}$ denotes the angle in which the particle rotates its front heading up or down. There is no third rotation axis used (``roll''), only ``pane'' (left-right) and ``tilt'' (up-down). \par
\medskip

In order to translate the motion from 2D to 3D the same motion law is applied twice, once for calculating the change in \boldmath $\varphi^{xy}$ and then also for the rotation of the tilt angle \boldmath $\varphi^{tilt}$. The following pseudocode describes the algorithm. The function ``in-cone'' reports all other agents in the 3D cone \underline{in front} of the particle, facing into the direction of its front heading, that is the direction it moves to with its forward motion. As the opening angle of the cone is $180$\textdegree, it is in fact a half-sphere. In order to get the particle in the right, left, up and down half-sphere, each individual particle always rotates into this direction, senses the number of particles in the half-sphere and then rotates back to its initial position. Given the fact that four half-spheres around the particle are processed, every neighboring particle is counted twice, once for the left-right rotational decision and another time for the up-down rotational decision. This gives the follow main loop for the simulation in 3D:
\begin{figure}[h!]
    \centering
    \includegraphics[width=0.8\textwidth]{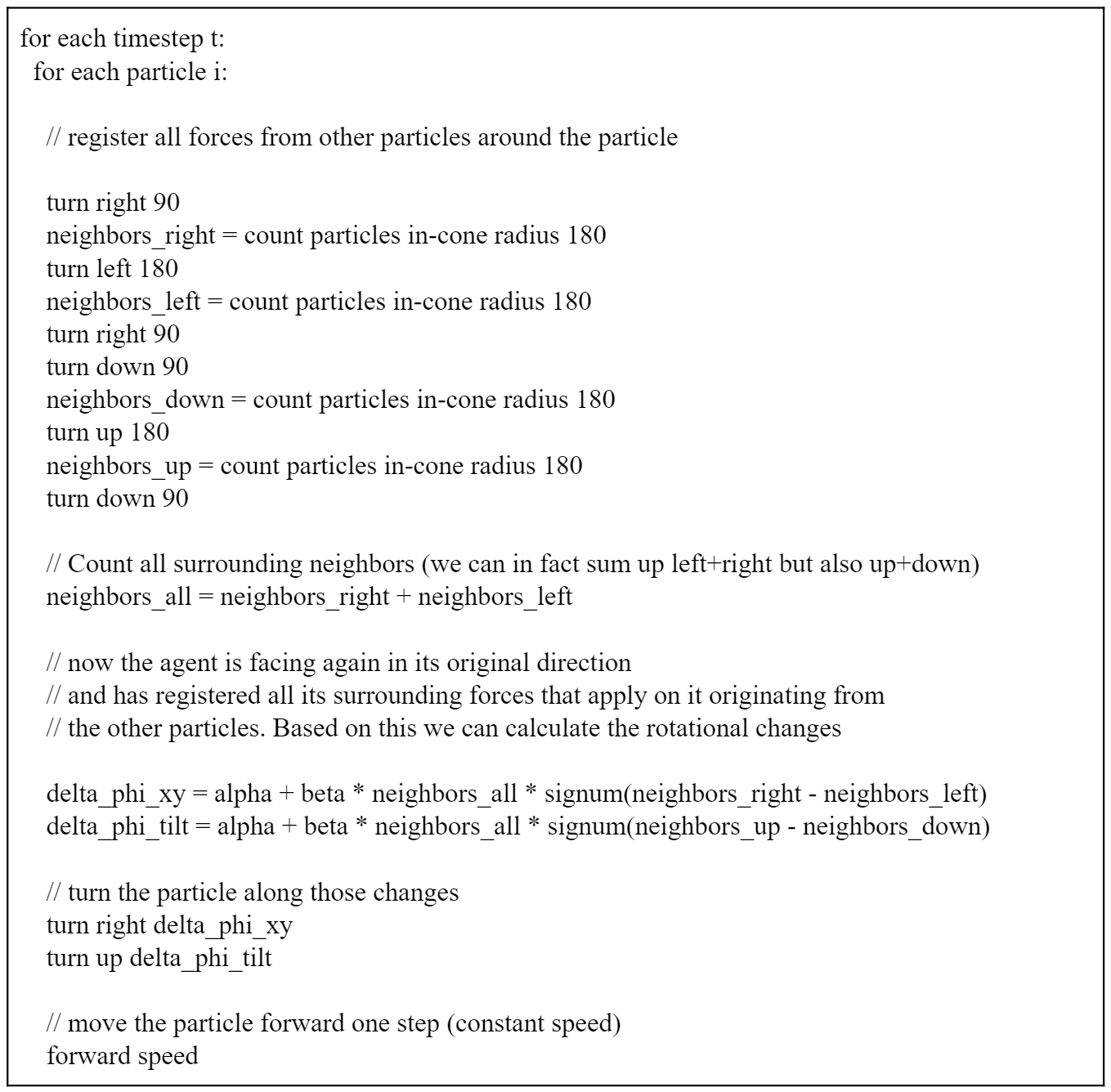}
    \label{fig:pseudo1}
\end{figure}
\medskip

(Please note that we use the signum function here, which returns $-1.0$ for all negative input numbers, $+1.0$ for all positive ones and $0.0$ for an input of zero. All angles are given in degrees and not in radians.)
Most parameters remain unchanged in the transition from the 2D to the 3D version. In order to allow the simulation to run faster (in 3D much more particles have to be calculated for the same density) we reduced the interaction radius to $r = 3.5$ space units. This reduced radius is then compensated by a slightly increased density-dependent rotation amount, thus $\beta = 24$ degrees in our 3D simulations. The fixed rotation $\alpha$ was kept on $180$\textdegree. The speed $v = 0.4$ space-units/step was also slightly decreased to fit to the reduced interaction radius. The density-dependent color scheme was adapted to fit the higher number of particles that one particle can encounter now:
\begin{figure}[h]
    \centering
    \includegraphics[width=0.8\textwidth]{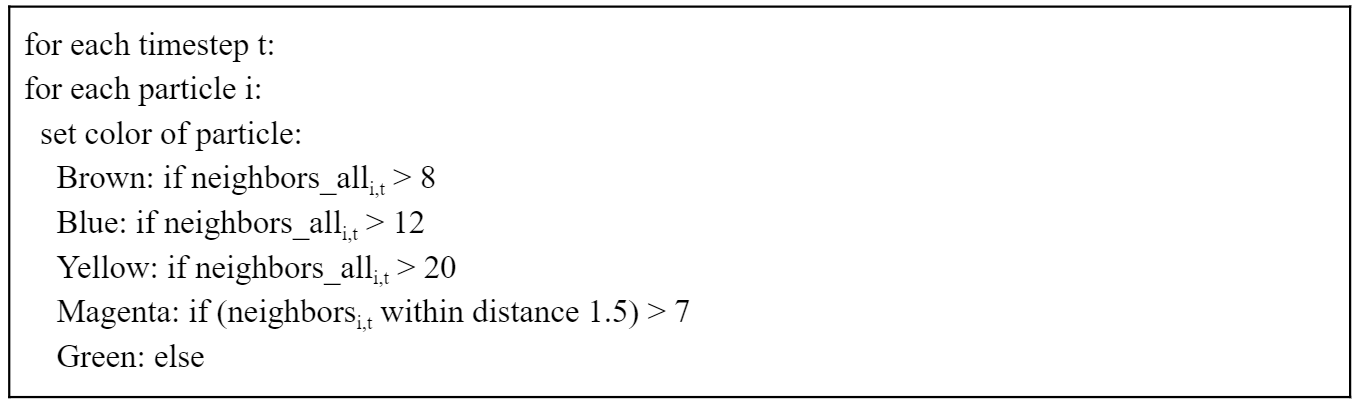}
    \label{fig:pseudo2}
\end{figure}

In order to allow to make the interesting denser (non-green) structures well visible and not being occluded by the free green particles we made all green particles semi-transparent.

\section{Results}The following Figure 1 exemplary shows an ecosystem of PPS structures that emerged from a random distribution:

\begin{figure}[ht]
\centering
   \begin{subfigure}{0.49\linewidth} \centering
     \includegraphics[scale=0.4]{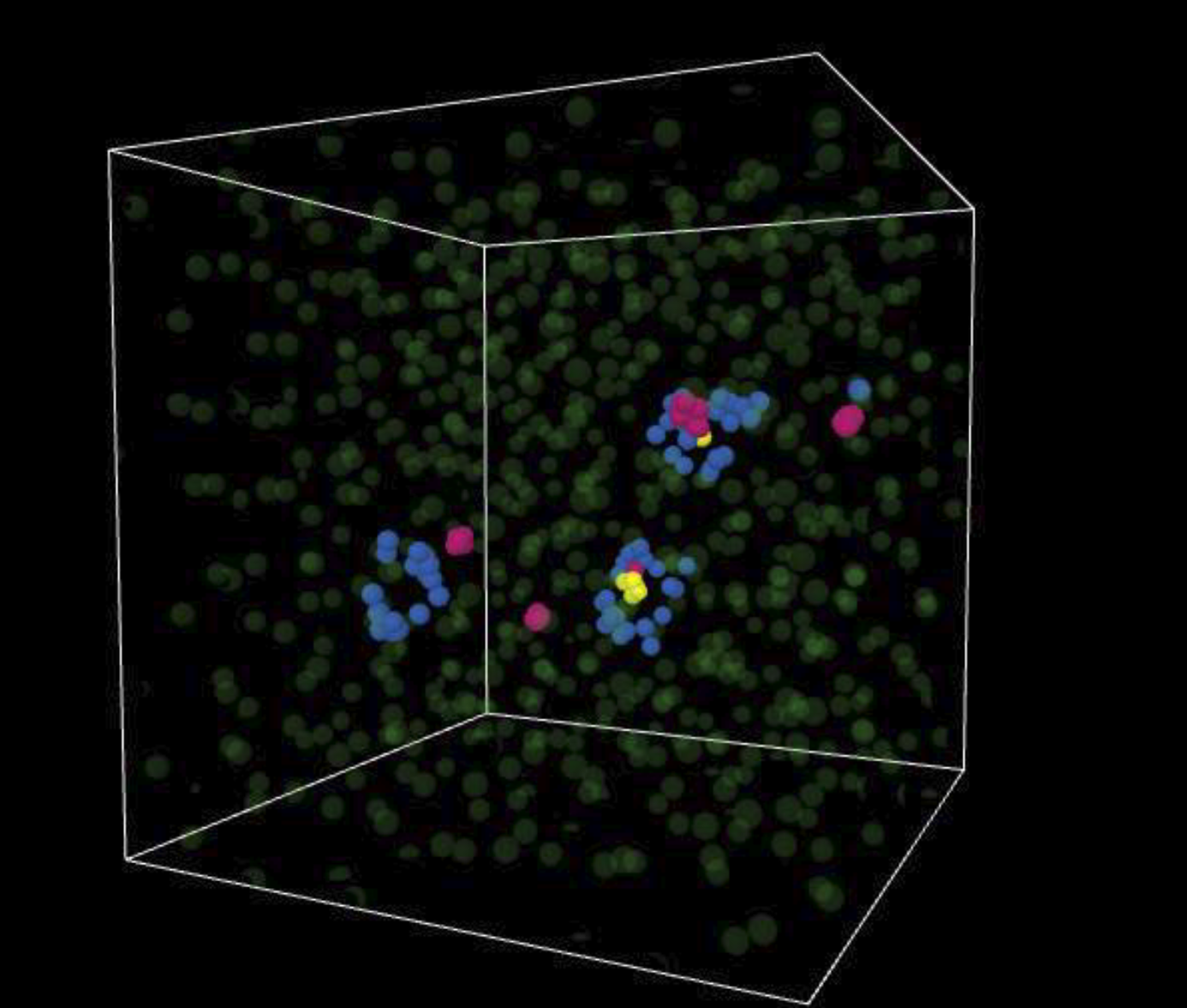}\label{fig:figA}
   \end{subfigure}
   \begin{subfigure}{0.49\linewidth} \centering
     \includegraphics[scale=0.395]{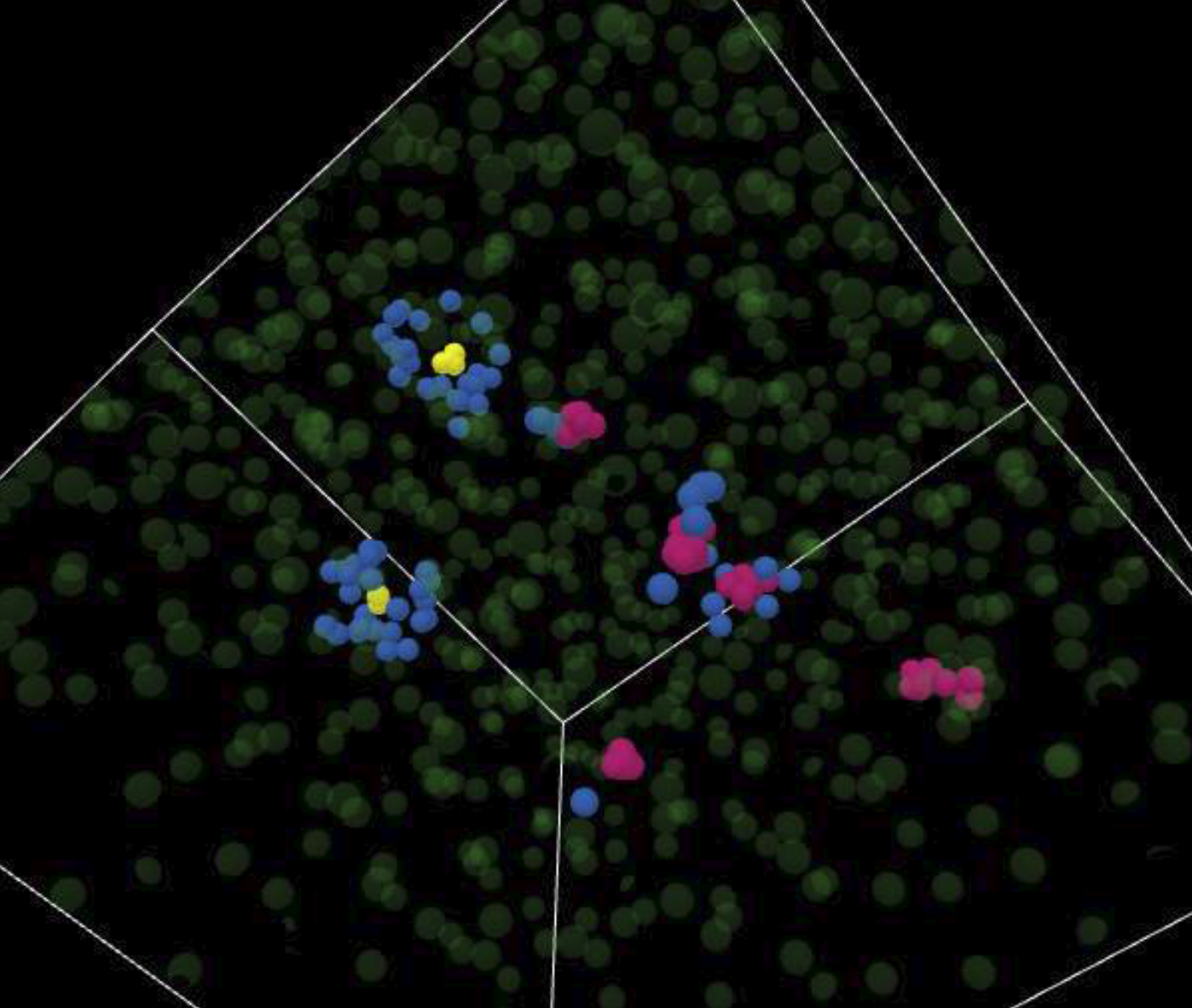}
    \label{fig:figB}
   \end{subfigure}
\caption{\textbf{Exemplary screenshots of the PPS in 3D.}} \label{fig:twofigs}
\end{figure}

\section{Discussion and Conclusion}

One of the first reactions to the model frequently expressed by other people is that it is very reminiscent of the Game of Life \cite{citation:4}. But not just the very different nature of the model (particles in continuous space that can move asynchronously) discriminates the two models. The fact that it is rather trivial to transform the Primordial Particle System to the third dimension, as shown here, indicates big differences. The changes to the original model are minimal, basically only the ``sensing'' of local interaction (forces) applied by neighboring particles is occurring now in half-spheres instead of half-circles as it was the case for the 2D version. The parameters actually did not require any change, we just reduced the radius $r$ by a bit and compensated for this by increasing $\beta$ and decreasing $v$ in order to let the system run at at feasible computational speed. This way fewer particle-to-particle interactions had to be accounted for but the achievable maximum rotation potential was kept on the same level as it was the case in the 2D version. The fact that parameters had to change only minimally (if at all necessary) and that the implementation is otherwise in fact identical to the 2D model just applied twice for 2 rotational axes, shows another important feature of the PPS model: it is simple and it stays simple even if extended to more dimensions.

\section{Acknowledgments}
This work was supported by the COLIBRI initiative of the University of Graz.

\clearpage 


\begin{thebibliography}{1}

\bibitem{citation:1}
Thomas Schmickl, Martin Stefanec, Karl Crailsheim: How a life-like system emerges from a simple particle motion law.
\textit{Scientific Reports}  {\bf 6}, 37969 (2016)., 
https://www.nature.com/articles/srep37969


\bibitem{citation:2} Thomas Schmickl, Martin Stefanec: Population Dynamics of Self-Replicating Cell-like Structures Emerging from Chaos. \textit{https://arxiv.org}, 1512.04478 (2015)., https://arxiv.org/abs/1512.04478

\bibitem{citation:3}
 Hiroki Sayama: Swarm chemistry.
\textit{Artificial life}  {\bf 15.1}, 105-114 (2009).

\bibitem{citation:4}
 Martin Gardner: Mathematical games: The fantastic combinations of John Conway\textquotesingle s new solitaire game ``life''. 
\textit{Scientific America}  {\bf 223}, 120-123 (1970).


\end{thebibliography}
\end{document}